\documentclass[reprint,superscriptaddress,amsmath,amssymb,aps,pra]{revtex4-1}

\usepackage{graphicx,epstopdf}
\usepackage{dcolumn}
\usepackage{bm}
\usepackage{hyperref}

\usepackage{multirow}
\usepackage{threeparttable}
\usepackage{color}
\usepackage{subfigure}
\usepackage{setspace}

\begin{document}
\title{Tomography-based Quantum Key Distribution}
\author{Yongtao Zhan}
\affiliation{Department of Electrical and Computer Engineering, University of Toronto, Toronto, Ontario, M5S 3G4, Canada}

\author{Hoi-Kwong Lo}
\email{hklo@ece.utoronto.ca}
\affiliation{Department of Electrical and Computer Engineering, University of Toronto, Toronto, Ontario, M5S 3G4, Canada}
\affiliation{Department of Physics, University of Hong Kong, Pokfulam, Hong Kong}

\date{\today}

\begin{abstract}
It has previously been shown that quantum state tomography can be used to increase the key rate of quantum key distribution (QKD) for the special case of qubits (i.e., $d=2$). Here, we consider the case of higher dimensions, i.e., qudits, and show that, for a prime number $d>2$, quantum state tomography can also improve the key rate of $d$-dimensional qudit-based QKD schemes, compared to the standard $(d+1)$-basis protocol. We apply our results to previous QKD experiment based on OAM (orbital angular momentum) encoding and demonstrate the advantage of tomography-based QKD protocols. Moreover, we compare the key rate of tomography-based QKD protocol with reference-frame-independent QKD protocol. We show that, for a rotation channel, the two protocols give the same key rate. However, for any other channels, tomography-based QKD protocol always gives a higher key rate than reference-frame-independent QKD protocol. 

\end{abstract}

\maketitle

\section{Introduction}
Quantum key distribution(QKD) offers a scheme for unconditionally secure communication. The most renown QKD protocols are BB84 protocol\cite{BB84} and six-state protocol\cite{PhysRevA.59.4238}. In these protocols, Alice prepare photon in conjugate basis and send it to Bob. Bob randomly select one of the conjugate basis to measure the received state. After that, they pick up the measurement results that they use the same basis, producing a shared random key. Although these protocols are secure theoretically, the gap between theory and experiment remains big. For example, quantum channel in real-life applications, usually is the optical fibre or free space, changes the quantum state of the photon during transmission. Many works has been done to improve the key generation rate of QKD in real-life channels. The tomography-based QKD protocol was proposed in Ref.\cite{PhysRevA.78.042316} to improve the key rate of QKD with accurate channel estimation. In BB84 and the six-state QKD protocols, data that are transmitted and received with mismatched bases are simply thrown away. For this reason, BB84 and the six-state QKD protocols only make use of the matched measurements $\{\sigma_i\sigma_i\}$ and consider diagonal matrix elements in the Bell-basis($\sigma_i$s are the Pauli matrices).  In contrast, a tomography-based QKD protocol keeps all the measurements $\{\sigma_i\sigma_j\}$, thus allows Alice and Bob to work out of the off-diagonal elements too. With the full matrix elements, one can get a higher key rate with a tomography-based QKD protocol (when compared to BB84 and the six-state protocol). Here in this paper, we extend the tomography-based QKD protocol to higher dimensions. We show how to perform quantum process tomography of the channel using the method introduced in Ref.\cite{PhysRevA.83.052332}. The tomography-based protocol also gives a higher key rate in high dimensional QKD compared to previous ones. The asymptotic key rate bound of the conventional $(d+1)$-basis protocol is given in Ref.\cite{PhysRevA.82.030301}, while we show tomography-based protocol can achieve an improvement over their result. By numerical simulating the two protocols in amplitude damping channel, we find the tomography-based protocol can tolerate higher noise. We apply our results to previous QKD experimental data and demonstrate the advantage of tomography-based QKD protocols. Hence tomography-based protocol can be utilized to improve the key rate in future real-life QKD applications.

In Ref.\cite{PhysRevA.82.012304}, Anthony et. al. proposed the so-called reference-frame-independent(RFI) QKD protocol. Their paper aims to address the problem that the reference frame of Alice(the sender) and Bob(the receiver) may not be well aligned, e.g. the polarization state may change in satellite-to-ground quantum communication due to the rotation of satellite. They showed how to bound the eavesdropper's information using Alice and Bob's measurement results. In this paper, we briefly summarize the two types of QKD protocols and compare them in different quantum channels. We show that the key rate of tomography-based QKD protocol is the same as RFI-QKD protocol in rotation channel. However, for any other channels including realistic optical fibre, the tomography-based QKD protocol gives a higher key rate. We come to the conclusion that the tomography-based protocol is a better choice than RFI-QKD in real-life QKD applications.

This paper is organized as follows. In Section.II, we will introduce the tomography-based QKD protocol and present a key rate formula for tomography-based QKD in the qubit case. In Section.III, we extend the tomography-based QKD to high-dimensional case, i.e. qudits, and show its advantage compared to the traditional $(d+1)$-basis protocol. Then we apply the high-dimensional tomography-based QKD to experiment using previous experimental data in Section.IV. The reference-frame-independent QKD protocol and high-dimensional RFI-QKD are introduced in Section.V and Section.VI. Finally we compare the RFI-QKD and tomography-based QKD in Section.VII.

\section{Key rate in Tomography-based QKD protocol}
Tomography-based Quantum Key Distribution\cite{PhysRevA.82.012304} can increase key rate with accurate channel estimation. In this section, we briefly review the protocol of tomography-based QKD and calculate the key rate of the tomography-based six-state protocol. In six-state protocol, Alice randomly sends the bit 0 or 1 to Bob by modulating it into a transmission basis that is randomly chosen from the z basis$({|0_z\rangle,|1_z\rangle})$, the x basis$({|0_x\rangle,|1_x\rangle})$, or the y basis$({|0_y\rangle,|1_y\rangle})$. In 
the entanglement-based version, Alice prepares a maximally entangled state $|\phi^+\rangle=\frac{1}{\sqrt{2}}(|0\rangle|0\rangle+|1\rangle|1\rangle)$ and sends the second qubit to Bob. The qubit channel $\mathcal{E}_B$ between Alice and Bob can be described by the affine map as follows \cite{PhysRevA.82.012304,NielsonChuang,PhysRevA.59.3290}
\begin{equation}
\left[\begin{array}{l}{\theta_{z}} \\ {\theta_{x}} \\ {\theta_{y}}\end{array}\right] \mapsto\left[\begin{array}{lll}{R_{z z}} & {R_{z x}} & {R_{z y}} \\ {R_{x z}} & {R_{x x}} & {R_{x y}} \\ {R_{y z}} & {R_{y x}} & {R_{y y}}\end{array}\right]\left[\begin{array}{l}{\theta_{z}} \\ {\theta_{x}} \\ {\theta_{y}}\end{array}\right]+\left[\begin{array}{l}{t_{z}} \\ {t_{x}} \\ {t_{y}}\end{array}\right]
\label{11}
\end{equation}    
where $(\theta_z,\theta_x,\theta_y)$ describes a vector in the Bloch sphere. This vector represents the state of a single qubit $\rho=\frac{1}{2}(I+\theta_{x}\sigma_{x}+\theta_{y}\sigma_{y}+\theta_{z}\sigma_{z})$ .

For the channel $\mathcal{E}_B$ and each pair of bases $(\mathrm{a}, \mathrm{b}) \in\{\mathrm{z}, \mathrm{x}, \mathrm{y}\}^{2}$, define the biases of the outputs as
\begin{equation} 
\begin{split}
Q_{ab0}&=\left\langle 0_{\mathrm{b}}\left|\mathcal{E}_{B}\left(\left|0_{\mathrm{a}}\right\rangle\left\langle 0_{\mathrm{a}}\right|\right)\right| 0_{\mathrm{b}}\right\rangle-\left\langle 1_{\mathrm{b}}\left|\mathcal{E}_{B}\left(\left|0_{\mathrm{a}}\right\rangle\left\langle 0_{\mathrm{a}}\right|\right)\right| 1_{\mathrm{b}}\right\rangle \\ 
Q_{ab1}&=\left\langle 1_{\mathrm{b}}\left|\mathcal{E}_{B}\left(\left|1_{\mathrm{a}}\right\rangle\left\langle 1_{\mathrm{a}}\right|\right)\right| 1_{\mathrm{b}}\right\rangle-\left\langle 0_{\mathrm{b}}\left|\mathcal{E}_{B}\left(\left|1_{\mathrm{a}}\right\rangle\left\langle 1_{\mathrm{a}}\right|\right)\right| 0_{\mathrm{b}}\right\rangle
\end{split}
\end{equation}
then the parameters of the channel in Eq.(\ref{11}) can be written as
\begin{equation}
    R_{ba}=\frac{1}{2}(Q_{ab0}+Q_{ab1}), t_b=\frac{1}{2}(Q_{ab0}-Q_{ab1})
\end{equation}

After Alice send the qubit to Bob, their joint state $\rho_{AB}$ can be written as:  

$\rho_{AB}=(id\otimes \mathcal{E}_B)|\phi^+\rangle \langle \phi^+| (id\otimes \mathcal{E}_B^{\dagger})=$
\begin{widetext}

\begin{equation}
\frac{1}{4}\left[\begin{array}{cccc}{1+R_{z z}+t_{z}} & {R_{x z}+t_{x}+\mathbf{i} R_{y z}+\mathbf{i} t_{y}} & {R_{z x}-\mathbf{i} R_{z y}} & {R_{x x}+R_{y y}+i R_{y x}-\mathbf{i} R_{x y}} \\ {R_{x z}+t_{x}-\mathbf{i} R_{y z}-\mathbf{i} t_{y}} & {1-R_{z z}-t_{z}} & {R_{x x}-R_{y y}-\mathbf{i} R_{y x}-\mathbf{i} R_{x y}}&{-R_{z x}+i R_{z y}} \\ {R_{z x}+\mathbf{i} R_{z y}} & {R_{x x}-R_{y y}+\mathbf{i} R_{y x}+\mathbf{i} R_{x z}} & {1-R_{zz}+t_z}&{-R_{x y}+t_{x}-\mathbf{i} R_{y z}+\mathbf{i}t_{y}} \\ {R_{x x}+R_{y y}-\mathbf{i} R_{y x}+\mathbf{i} R_{x y}} & {-R_{z x}-\mathbf{i} R_{z y}} & {-R_{x z}+t_{x}+\mathbf{i} R_{y z}-\mathbf{i}t_{y}} & {1+R_{z z}-t_{z}}\end{array}\right]
\label{22}
\end{equation}
\end{widetext}

That means, by analyzing the full probability distribution of their measurement results, Alice and Bob can have the full information of the channel. We treat only Alice's bit transmitted in $z$ basis and corresponding Bob's bit sequence received in $\sigma_z$ measurement. In this tomography-based QKD protocol, the key rate of direct reconciliation is \cite{dw}
\begin{equation}
r_{QST}=I_{\rho}(X|Y)-\chi_{\rho}(X:E)
\label{55}
\end{equation}
where $I_{\rho}(X|Y)$ is the mutual information between Alice and Bob, and $\chi_{\rho}(X:E)$ denotes the Holevo quantity between Alice and Eve.
Here we give an explicit form of the Devetak-Winter rate in Eq.(\ref{55}) as an function of $\rho_{AB}$:
\begin{equation} 
I_{\rho_{AB}}(X|Y)=1+\sum_{i=1}^{4} \delta_i log_2\delta_i+h(\delta_1+\delta_3)
\label{66}
\end{equation}
\begin{equation} 
\chi_{\rho_{AB}}(X:E)=S(\rho_{AB})-\frac{1}{2}S(_A\langle 0|\rho_{AB}|0 \rangle_A)-\frac{1}{2}S(_A\langle 1|\rho_{AB}|1 \rangle_A)
\end{equation}
\begin{equation} 
\begin{split}
    r_{QST}=&I_{\rho_{AB}}(X|Y)-\chi_{\rho_{AB}}(X:E)\\
    =&1+\sum_{i=1}^{4} \delta_i log_2\delta_i+h(\delta_1+\delta_3) \\
    &-S(\rho_{AB})+\frac{1}{2}S(_A\langle 0|\rho_{AB}|0 \rangle_A)+\frac{1}{2}S(_A\langle 1|\rho_{AB}|1 \rangle_A)
\end{split}
\label{88}
\end{equation}

 where $\delta_i$s are the diagonal elements of the density matrix in the computational basis.
 \begin{equation}
     \begin{split}
         \delta_1&=\langle 00|\rho_{AB}|00 \rangle, \delta_2=\langle 01|\rho_{AB}|01 \rangle, \\
         \delta_3&=\langle 10|\rho_{AB}|10 \rangle, \delta_4=\langle 11|\rho_{AB}|11 \rangle
     \end{split}
 \end{equation}

We plug the density matrix in Eq.(\ref{22}) into Eq.(\ref{88}) and get the key rate in tomography-based QKD:
\begin{equation} 
\begin{split}
    r_{QST}=& 1-S(\rho_{AB})+\sum_{i=1}^{4} \delta_i log_2\delta_i+h(\frac{1+t_z}{2}) \\
    &+\frac{1}{2}h((1+\sqrt{(R_{zz}+t_z)^2+(R_{xz}+t_x)^2+(R_{yz}+t_y)^2})/2)
    \\
    &+\frac{1}{2}h((1+\sqrt{(R_{zz}-t_z)^2+(R_{xz}-t_x)^2+(R_{yz}-t_y)^2})/2)
\end{split}
\end{equation}    
\section{Extending Tomography-based QKD protocol to high dimensions}
\subsection{Security of high dimensional QKD}
Traditional QKD protocols are based on two level systems, or qubits. Some QKD protocols using higher-dimensional quantum systems(or qudits) have been proposed. These qudit-based protocols can achieve higher key rates and security tolerance to noise\cite{PhysRevLett.88.127902}.

There are two main families of high dimensional QKD protocols.  The first protocol is $(d+1)$-basis protocol, which is the generalization of six-state protocol. In this protocol, Alice and Bob choose one of $(d+1)$ mutually unbiased basis(MUB) in $d$-dimensional Hilbert space to send and measure the state independently. Then they compare the basis they use and keep the measurement results that they use the same basis. The second is two-basis protocol, which is the generalization of BB84 protocol. It is similar to the $(d+1)$-basis protocol except for Alice and Bob using two MUBs to encode the information instead of $(d+1)$ MUBs.  For simplicity we only focus on the $(d+1)$-basis protocol here.

Here we briefly introduce the concept of MUB. Two orthonormal basis $\mathcal{M}_{1}=\left\{\left|\phi_{(1, i)}\right\rangle, i=0,1, \ldots, d-1\right\}$ and  $\mathcal{M}_{2}=\left\{\left|\phi_{(2, j)}\right\rangle, j=0,1, \ldots, d-1\right\}$ of a $d$-dimensional Hilbert space $\mathcal{H}_d$ are said to be mutually unbiased if and only if all pairs of basis vector satisfy
\begin{equation}\left|\left\langle\phi_{(1, i)} | \phi_{(2, j)}\right\rangle\right|^{2}=\frac{1}{d}\end{equation}

Physically, this means if a state in $\mathcal{M}_1$ is measured with respect to $\mathcal{M}_2$, all outcomes are equal probable.  This property makes them important for QKD protocols. Wootters and Fields\cite{annal} proved that the number of MUBs one may find for any dimension $d$ is at most $(d+1)$ and the $(d+1)$ MUBs do exist whenever $d$ is a power of a prime. So our discussion is restricted to the cases that the dimension $d$ is a prime power.

The security of high dimensional QKD based on MUBs is analyzed in Ref.\cite{PhysRevA.82.030301}. Here we briefly summarize their results. The MUBs in their  QKD protocol are chosen to be eigenbasis of Weyl operators $U_{jk}$. The Weyl operators are generalization of Pauli operators and are defined by $U_{j k}=\sum_{s=0}^{d-1} \omega^{s k}|s+j\rangle\langle s|$  for $j,k\in\{0,1,2,...,d-1\}$ and $\omega$ is the $d$-th root of unity. The choice of $(d+1)$ MUBs are the eigenbasis of  $\{U_{01},U_{1k}:k\in[0,d-1]\}$. In the $(d+1)$-basis protocol, Alice and Bob choose one of the $(d+1)$ MUBs to send and measure the state. After they get their raw keys, they can estimate the error vectors
\begin{equation}\underline{q_{j k}}=\left\{q_{j k}^{(0)}, q_{j k}^{(1)}, \ldots, q_{j k}^{(d-1)}\right\}\end{equation}
where $q_{j k}^{(t)}=\operatorname{Prob}(a-b=t \bmod d | j, k)$ is the probability that Alice's outcome $a$ and Bob's outcome $b$ differed by $t$, modulo by $d$, when the basis of $U_{jk}$ is chosen by both. We note that $q^{(0)}_{jk}$ is the probability that Alice and Bob's measurement result matches.

They construct the density matrix of Alice and Bob's joint state  with the estimated error vectors. The density matrix is diagonal in the generalized Bell basis\cite{PhysRevA.82.030301}.
\begin{equation}
\rho'=\sum_{j,k=0}^{d-1}\lambda_{jk}|\Phi_{jk} \rangle \langle \Phi_{jk}|
\label{4444}
\end{equation}
The generalized Bell basis are given by $\left|\Phi_{j k}\right\rangle=\sum_{s=0}^{d-1} \omega^{s k}|s \;s+j\rangle=\mathbb{I} \otimes U_{j k} | \Phi_{00}\rangle$. The eigenvalues $\lambda_{jk}$ are given in Eq.(4) in Ref.\cite{PhysRevA.82.030301}, expressed by the estimated error vectors

\begin{equation}
\lambda_{j k}=\frac{1}{d}\left(\sum_{s} q_{1 s}^{(s j-k \bmod d)}+q_{01}^{(j)}-1\right)
\end{equation}

Although the Bell diagonal mixture may not be the real joint state, it produces the same statistics for Alice's and Bob's measurements. We can suppose without loss of generality that Alice and Bob rather receive the state described in Eq.(\ref{4444}).

The asymptotic key rate of $(d+1)$-basis QKD protocol is also given in Ref.\cite{PhysRevA.82.030301}. For asymptotic bounds, We can assume that only one basis is used for generate keys and is chosen almost always, while the other basis are chosen with negligible probability and used to bound the eavesdropper's information. In Ref.\cite{PhysRevA.82.030301}, They choose the key basis to be the one of $U_{01}$. The Devetak-Winter rate is 
\begin{equation}
r_{\infty}=I(A:B)-\chi(A:E)
\end{equation}
We can easily get the mutual information is $I(A:B)=log_2d-H(\underline{q_{01}})$. In Ref.\cite{PhysRevA.82.030301} they get Eve's information
\begin{equation}
\chi(A: E)=H(\underline{\lambda})-H(\underline{q_{01}})
\end{equation}
where $\underline{\lambda}=\{\lambda_{jk}\}$ for $ j,k\in[0,d-1]$ are the eigenvalues of the density matrix $\rho_{AB}$.

\subsection{High Dimensional Tomography-based QKD}

In entanglement-based high dimensional QKD protocol, Alice prepare a $d$-dimensional entangled state $|\Phi_{00}\rangle=\frac{1}{\sqrt{d}}\sum_{i=0}^{d-1} |i\rangle_A |i\rangle_B$ and send one of the qudit to Bob. After Bob receive the qudit, they randomly choose one of $(d+1)$ mutual unbiased projectors(MUB-projectors) to measure the state. The set of MUB-projectors is given by
\begin{equation}\mathcal{P}_{m}^{(\gamma)}=\left|\psi_{m}^{(\gamma)}\right\rangle\left\langle\psi_{m}^{(\gamma)}\right|, m=1, \ldots, d, \gamma=0, \ldots, d\end{equation}
where $\gamma$ labels one of the $(d+1)$ families of MUBs and $m$ denotes one of the orthogonal state in this family. Instead of only keeping the measurement results that they use the same basis, they use all the results to perform quantum channel tomography.

Here we use the method introduced in Ref.\cite{PhysRevA.83.052332} to perform quantum process reconstruction. We consider the quantum channel as a general evolution of the qudit described by a completely positive linear map:
\begin{equation}
    \mathcal{E}(\rho)=\sum_{i} A_{i} \rho A_{i}^{\dagger}\label{4848}
\end{equation}
Here $\rho$ is the state that Alice sends to Bob and $A_i$s are the Kraus operators. We use the overcomplete basis of $(d^2+d)$ MUB-projectors to expand the Kraus operators:

\begin{equation}A_{i}=\sum_{\alpha=0}^{d} \sum_{m=1}^{d} a_{i m}^{(\alpha)} \mathcal{P}_{m}^{(\alpha)}\end{equation}
Then the complete linear map $\mathcal{E}(\rho)$ of the quantum channel can be expressed in the following manner:

\begin{equation}\mathcal{E}(\rho)=\sum_{\alpha, \beta=0}^{d} \sum_{m, n=1}^{d} \chi_{m n}^{(\alpha, \beta)} \mathcal{P}_{m}^{(\alpha)} \rho \mathcal{P}_{n}^{(\beta)}\end{equation}
where $\chi_{m n}^{(\alpha, \beta)} \equiv \sum_{i} a_{i m}^{(\alpha)} a_{i n}^{(\beta) *}$ is the process matrix. This equation is Eq.(6) in Ref.\cite{PhysRevA.83.052332}. We can do full tomography of the quantum channel by determining all elements of the process matrix. Next we show how Alice and Bob perform quantum channel tomography with their measurement results.

We consider the full probability distribution of Alice's and Bob's data. The probability of Alice and Bob choosing each family of MUBs is $1/(d+1)$. If Alice choose the $l$-th family and Bob choose the $s$-th family, the probability of Alice getting $\left|\psi_{l}^{(\gamma)}\right\rangle$ and Bob getting $\left|\psi_{s}^{(\eta)}\right\rangle$  is 

\begin{equation}
\begin{aligned}
p_{\eta s}^{(\gamma, l)}&=Tr(\mathcal{E}(\rho)\mathcal{P}_{s}^{(\eta)})\\
&=\sum_{\alpha, \beta=0}^{d} \sum_{m, n=1}^{d} \chi_{m n}^{(\alpha, \beta)} \operatorname{Tr}\left(\mathcal{P}_{m}^{(\alpha)} \mathcal{P}_{l}^{(\gamma)} \mathcal{P}_{n}^{(\beta)} \mathcal{P}_{s}^{(\eta)}\right)
\end{aligned}
\label{5151}
\end{equation}
This equation is Eq.(11) in Ref.\cite{PhysRevA.83.052332}. Alice and Bob can provide a good estimation of $p_{\eta s}^{(\gamma, l)}$ using their measurement results. There are $d^2(d+1)^2$ unknown elements $\chi_{m n}^{(\alpha, \beta)}$ and $d^2(d+1)^2$ linear equations in total. By solving these linear equations, they can reconstruct the quantum channel.

After Alice and Bob reconstruct the quantum channel, they can find all the elements of the density matrix of their joint state $\rho_{AB}$.
\begin{equation}
    \rho=\sum_{j,k,m,n=0}^{d-1}\rho_{jk}^{mn} |\Phi_{jk} \rangle \langle \Phi_{mn}|
\end{equation}
The matrix elements $\rho_{jk}^{mn}$ are given by $\rho_{jk}^{mn}=\sum_i \langle \Phi_{jk}|(\mathbb{I} \otimes A_i)|\Phi_{00}\rangle\langle\Phi_{00}|(\mathbb{I} \otimes A_i^\dagger)|\Phi_{mn}\rangle$.
They can estimate Eve's information and perform postprocessing of their raw key in a similar way as the original tomography-based protocol\cite{PhysRevA.78.042316}.

\subsection{Asymptotic key rate of High Dimensional Tomography-based QKD}
Alice and Bob can reconstruct the quantum state $\rho$ they share through quantum channel tomography. Similar to the two dimensional case in Section.II, we can calculate the asymptotic key rate of high dimensional tomography-based QKD. The classical mutual information and the Holevo quantity are
\begin{equation} 
I_{\rho}=log_2d+\sum_{i,j=0}^{d-1} \delta_{ij} log_2\delta_{ij}+H(\underline{\delta})
\end{equation}
\begin{equation} 
\chi_{\rho}=S(\rho)-\frac{1}{d}\sum_{i=0}^{d-1}S(_A\langle i|\rho|i \rangle_A)
\label{5555}
\end{equation}
where $\delta_{ij}$ are the diagonal elements of $\rho$ in the computational basis
\begin{equation}
    \delta_{ij}=\langle ij|\rho|ij \rangle
\end{equation}
and $\underline{\delta}=\{\sum_{j=1}^{d} \delta_{ij} \}$ for $i\in[0,d-1]$. Once we get the mutual information and the Holevo quantity, we can calculate the key rate using the Devetak-Winter rate formula.

\subsection{Comparison of tomography-based protocol and $(d+1)$-basis protocol}
In conventional $(d+1)$-basis protocol, Alice and Bob can find out the diagonal elements of the density matrix in generalized Bell basis.  However, in tomography-based protocol, they can perform quantum channel tomography and find out all the elements of the density matrix. With the full information of their quantum state, they can make a better estimation of Eve's information compared to conventional $(d+1)$-basis protocol, thus improving the key rate.

In tomography-based QKD protocol, Eve's information is quantified by the Holevo quantity $\chi_{\rho}$, which is given in Eq.(\ref{5555}) in previous section. In $(d+1)$-basis protocol, Eve's information is

\begin{equation}
    \chi_{\rho'}=S(\rho')-\frac{1}{d}\sum_{i=0}^{d-1}S(_A\langle i|\rho'|i \rangle_A)
\end{equation}

We can prove that
\begin{equation}
    \chi_{\rho}\leq\chi_{\rho'}
\end{equation}

The classical mutual information $I(A:B)$ are the same for both protocols. So we come to the conclusion that the tomography-based protocol gives a higher key rate than the $(d+1)$-basis protocol.

 Here we compare the tomography-based protocol and conventional $(d+1)$-basis protocol in three dimensional amplitude damping channel as an example. The amplitude damping channel is a model describing the decay process of multi-level atoms\cite{8248788}. It can be described by a completely positive linear map in Eq.(\ref{4848}). The superoperators are\cite{8248788}
\begin{equation}
\begin{array}
{l}
A_{0}=|0\rangle\langle 0|+\sqrt{1-\alpha}| 1\rangle\langle 1|+(1-\alpha)| 2\rangle\langle 2| \\
A_{1}=\sqrt{\alpha}|0\rangle\langle 1|+\sqrt{2 \alpha(1-\alpha)}| 1\rangle\langle 2| \\
A_{2}=\alpha|0\rangle\langle 2|
\end{array}
\end{equation}

parametrized by a real parameter $0\leq \alpha \leq 1$.

We calculate the asymptotic key rates for tomography-based protocol and conventional $(d+1)$-basis protocol using the schemes described in Section.III.A and Section.III.C. The explicit formulas are very complicate so I don't present them here. The key rates of both protocols are plotted in Figure.1. We find the key rate of  tomography-based protocol is always higher than the conventional $(d+1)$-basis protocol. 

\begin{figure}
\includegraphics[width=0.5 \textwidth]{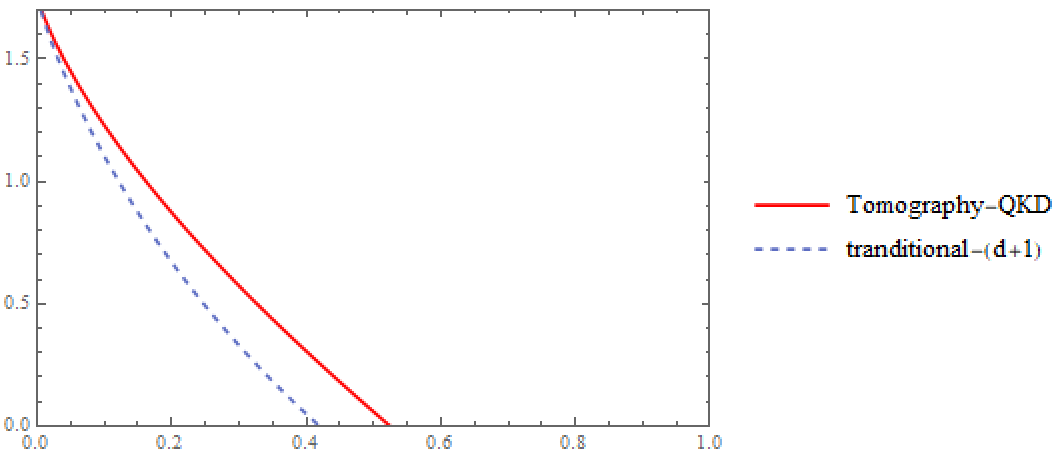}
\caption{The key rate of high dimensional tomography-based protocol and traditional $(d+1)$-basis protocol as a function of $\alpha$ in qutrit amplitude damping channel. The horizontal coordinate is $\alpha$ and the vertical coordinate is the key rate. We find the key rate of tomography-based protocol reach zero when $\alpha\approx0.5$. For traditional $(d+1)$-basis protocol, the value is about 0.4. The tomography-based protocol gives a higher key rate and is more robust to noise.} 
\label{fig-1}
\centering
\end{figure}

\section{High Dimensional Tomography-based QKD in Experiment}
There are multiple experimental implementation of RFI-QKD\cite{Wabnig_2013,sc,PhysRevA.95.032309,PhysRevLett.112.130501}. However, to our knowledge, the tomography-based QKD protocol hasn't been implemented yet. Here we try to demonstrate the high dimensional tomography-based QKD in experiment.

In Ref.\cite{PhysRevA.88.032305} the authors present an experimental study of higher-dimensional quantum key distribution protocols based on mutually unbiased bases. The qudits are realized by the orbital angular momentum state of photons. They perform $(d + 1)$ mutually unbiased measurements on a pair of entangled photons for dimensions ranging from $d = $2 to 5. For simplicity we only focus on $d=3$ case in this paper.  In the high dimensional QKD protocol, the probability of Alice and Bob choosing each family of MUBs is $1/(d+1)$. If Alice choose the $l$-th family and Bob choose the $s$-th family, the probability of Alice getting $\left|\psi_{l}^{(\gamma)}\right\rangle$ and Bob getting $\left|\psi_{s}^{(\eta)}\right\rangle$  is $p_{\eta s}^{(\gamma, l)}$ in Eq.(\ref{5151}). In $d=3$ case, the full probability distribution of Alice's and Bob's data $p_{\eta s}^{(\gamma, l)}$ is a $12\times12$ matrix. They measured the matrix in experiment and their result is depicted in Figure.2. We note that the diagonal elements are approximately equal to 1/3 and the elements corresponding
to different bases are found to be approximately 1/9. That corresponds to the property of MUBs which we mentioned in Section.III.A .Figure.3 contains their measured secret key rate in the experiment. Then they made a comparison between their experimental data and theoretical key rate in the qudit depolarizing channel. We note that it is only a coincidence that the measured results seems fit precisely the theory curve in $d=3$ case. They did the QKD experiment in $d=2,3,4,5$ cases and the measured results fit not very well with the theory curve when $d=2,4,5$. 

In fact, we can apply the tomography-based protocol with the full probability distribution of Alice's and Bob's data. Using the schemes described in Section.III.A, we can perform quantum channel tomography based on MUBs. Based on the equations in Section.III.C, we then calculate the key rate of tomography-based protocol, which was contained in Figure.3. We find the tomography-based protocol yields a higher key rate than conventional $(d+1)$-basis protocol.
\begin{figure}
\centering
\includegraphics[width=0.45 \textwidth]{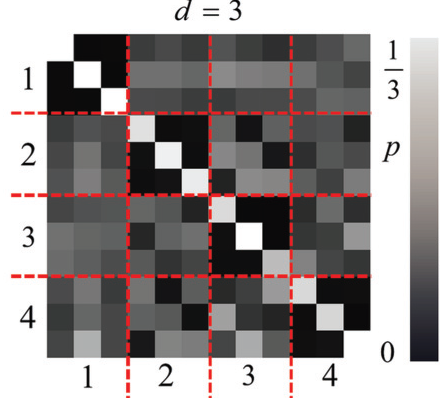}
\caption{The normalized joint probabilities of Alice and Bob's data. The numbers on the horizontal/vertical axis labels the MUB that Alice/Bob choose. In the entanglement-based scheme, Alice and Bob's measurement results are one of the three states in one of the four basis.} 
\label{fig-2}
\end{figure}
\begin{figure}
\centering
\includegraphics[width=0.5 \textwidth]{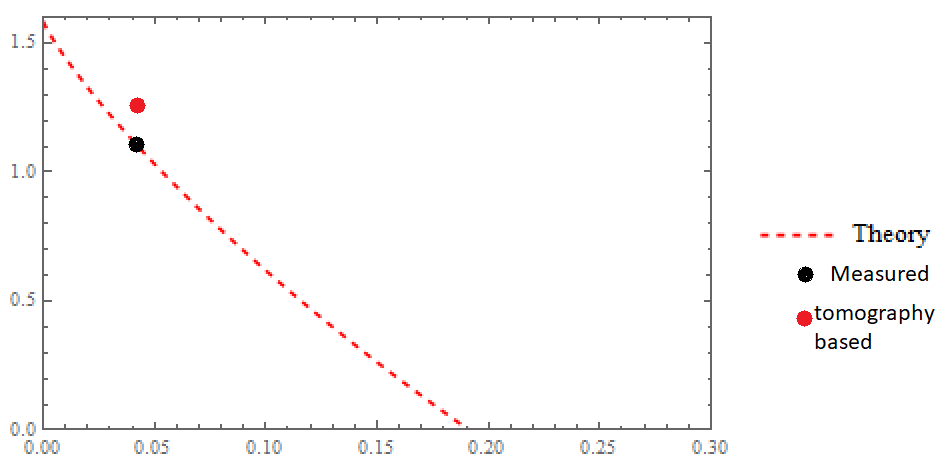}
\caption{The secret key rate as a function of the
average error rate in the qutrit depolarizing channel. The horizontal coordinate is the average error rate and the vertical coordinate is the key rate. The solid black data points
denote the measured values in Ref.\cite{PhysRevA.88.032305} and the dashed curves the theoretical values. The solid red data points is the key rate of the tomography-based protocol.}
\end{figure}

\section{Key rate in RFI-QKD protocol[3]}
In this section, we briefly review the RFI-QKD protocol which was introduced in Ref.\cite{PhysRevA.82.012304}. In the RFI-QKD protocol, we assume one of the three directions is well aligned and the other two directions are rotating dependent of time. Take satellite-to-ground QKD as an example. Vertical and horizontal polarization state may vary in time because of the rotation of the satellite but the circular state is stable. For the well-aligned direction we set $Z_A=Z_B$. The other two directions are related by $X_B=cos\beta X_A + sin\beta Y_A$ and $Y_B=cos\beta Y_A-sin\beta X_A$ and $\beta$ may change in time. We assume that the frames are varying very slowly that Alice and Bob can collect enough signals to create a key above the finite-size effect in a time interval that $\beta$ does not vary too much. In this protocol we consider the worst-case scenario that the two directions are fixed and known to Eve. In each run, Alice prepares a maximally entangled state $|\phi^+ \rangle$ and sends the second qubit to Bob. We assume that Alice and Bob only use the $Z$ basis to generate the key. So the quantum bit error rate(QBER) is given by
\begin{equation}
    Q=\frac{1-\langle Z_A Z_B\rangle}{2}
\end{equation}
Alice and Bob use another quantity to bound Eve's knowledge, which reads
\begin{equation}
    C=\langle X_A X_B\rangle^2+\langle X_A Y_B\rangle^2+\langle Y_A X_B\rangle^2+\langle Y_A Y_B\rangle^2
    \label{1212}
\end{equation}

We note that C is invariant under the transformation $X_A  \rightarrow -X_A$, $Y_A  \rightarrow -Y_A$, $X_B  \rightarrow -X_B$ and $Y_B  \rightarrow -Y_B$. In the presence of such a symmetry, we can replace $\rho_{AB}$ by $ \tilde{\rho}_{AB}=\frac{1}{2}(\rho_{AB}+Z_AZ_B\rho_{AB} Z_AZ_B)$, so we have
\begin{equation}
\begin{aligned} \tilde{\rho}_{A B}=& \mu_{1} P_{\Phi^{+}}+\mu_{2} P_{\Phi^{-}}+\left(\frac{a}{2}\left|\Phi^{-}\right\rangle\left\langle\Phi^{+}\right|+\mathrm{H.c.}\right) \\ &+\mu_{3} P_{\Psi^{+}}+\mu_{4} P_{\Psi^{-}}+\left(\frac{b}{2}\left|\Psi^{-}\right\rangle\left\langle\Psi^{+}\right|+\mathrm{H.c.}\right) \end{aligned}
\end{equation}
where $P_{\psi}=|\psi\rangle\langle\psi|$ and the four states represent the Bell basis. For convenience of notation, let us call this state $\rho(a,b)$, we find $C$ and $Q$ are the same for state $\rho(-a^*,-b^*)$. So by the same argument, we can study the mixture $\rho_{AB}'=\frac{1}{2}(\rho(a,b)+\rho(-a^*,-b^*))$.
This state is:
\begin{equation}
\begin{aligned} {\rho}_{A B}'=& \mu_{1} P_{\Phi^{+}}+\mu_{2} P_{\Phi^{-}}+\left(\frac{Im(a)}{2}\left|\Phi^{-}\right\rangle\left\langle\Phi^{+}\right|+\mathrm{H.c.}\right) \\ &+\mu_{3} P_{\Psi^{+}}+\mu_{4} P_{\Psi^{-}}+\left(\frac{Im(b)}{2}\left|\Psi^{-}\right\rangle\left\langle\Psi^{+}\right|+\mathrm{H.c.}\right) \end{aligned}
\end{equation}
This state can be written as Bell diagonal:
\begin{equation}
    \rho_{AB}'=\sum^{4}_{i=1}\lambda_i'|\Phi_i\rangle\langle\Phi_i|
\end{equation}
Since we only consider collective attacks, the key rate of RFI-QKD protocol can also be calculated using Eq.(\ref{55}), while the density matrix we are studying is $\rho_{AB}'$:
\begin{equation} 
\begin{split}
    r_{RFI}&=I_{\rho_{AB}'}(X|Y)-\chi_{\rho_{AB}'}(X:E)\\ &=1-S(\rho_{AB}')+\sum_{i=1}^{4} \delta_i' log_2\delta_i'+h(\delta_1'+\delta_3') \\
    &+\frac{1}{2}S(_A\langle 0|\rho_{AB}'|0 \rangle_A)+\frac{1}{2}S(_A\langle 1|\rho_{AB}'|1 \rangle_A)
\end{split}
\label{1616}
\end{equation}
where $\delta_i$s are the diagonal elements of $\rho_{AB}'$ in the computational basis. We find the latter four terms in Eq.(\ref{1616}) cancel out, so we have
\begin{equation}
    r_{RFI}=1-S(\rho_{AB}')
\end{equation}
The two protocols have the same formula of key rate but for different density matrix. In the RFI-QKD protocol, Alice and Bob have less information about their joint state, so intuitively we have
\begin{equation}\label{ineq}
    r_{QST}\geq r_{RFI}
\end{equation}
We will give a rigorous proof of this in the appendix.
\section{Extending RFI-QKD protocol to higher dimensions}

The original RFI-QKD paper\cite{PhysRevA.82.012304} also discussed high dimensional RFI-QKD protocols. I didn't make any meaningful progress in this topic, so I just write a brief summary of their results here.

In qubit case, we use the constant $C$ in Eq.(\ref{1212}) to quantify the entanglement and bound Eve's knowledge. Similarly, we can calculate a phase-invariant constant $C_d$ using our measurement result in high dimensional case.

For example, in three-dimensional case, they find
\begin{equation}C_{3}=\sum_{i=2}^{4} \sum_{j=2}^{4} e_{i j} e_{i j}^{*}+\sum_{i=2}^{4} \sum_{j=-2}^{-4} e_{i j} e_{i j}^{*} \leqslant 3\label{6060}\end{equation}
where $e_{ij}$ is defined by \begin{equation}
    e_{i j}=\operatorname{Tr}\left(\tau_{i} \otimes \tau_{j} \rho_{A B}\right)
\end{equation}
and $\tau_i$ is the $i$-th so-called Weyl operators. The maximum value of $C_3$ is 3, achievable only by two qutrit maximally entangled states.
In order to calculate the lower bound of the key rate, we have to consider the worst case and do the optimization. However, there is no analytical optimization for high dimensional RFI-QKD protocols.

In practical cases, we recommend the tomography-based protocol rather than the RFI-QKD protocol because we still don't know how to measure the phase invariant $C_d$ in Eq.(\ref{6060}) in experiment because it contains expectation value of non-Hermitian operators and $e_{ij}$ may not be real, which makes the high dimensional RFI-QKD protocol difficult to implement. The tomography-based protocol not only gives a higher key rate but also has an analytical key rate formula. However, the tomography-based protocol has defect when compared to the conventional $(d+1)$-basis protocol. Although it gives a higher key rate, it requires more measurements to perform precise channel tomography. It is better to choose the $(d+1)$-basis protocol when we do not have enough measurement results. For future high-rate QKD applications, we should use the tomography-based protocol instead.
\section{Comparison of key rates in different channels}
\subsection{Amplitude damping channel}
The key rates of the above two protocols are multivariable functions which makes it difficult to make a direct comparison. We consider applying the two protocols in amplitude damping channel.
This channel $\mathcal{E}_p$ is given by the affine map parametrized by a real parameter $0\leq p \leq 1$.
\begin{equation}
\left[\begin{array}{l}{\theta_{z}} \\ {\theta_{x}} \\ {\theta_{y}}\end{array}\right] \mapsto\left[\begin{array}{ccc}{1-p} & {0} & {0} \\ {0} & {\sqrt{1-p}} & {0} \\ {0} & {0} & {\sqrt{1-p}}\end{array}\right]\left[\begin{array}{l}{\theta_{z}} \\ {\theta_{x}} \\ {\theta_{y}}\end{array}\right]+\left[\begin{array}{l}{p} \\ {0} \\ {0}\end{array}\right]
\label{1919}
\end{equation}
\begin{figure}
\includegraphics[width=0.5\textwidth]{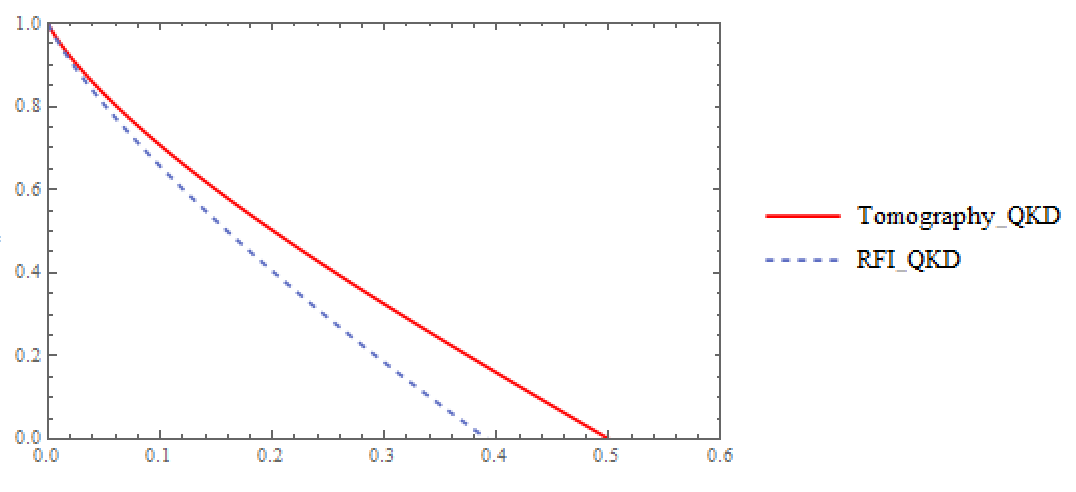}
\caption{The key rate of Tomography-based QKD and RFI-QKD in amplitude damping channel. The horizontal coordinate is $p$ and the vertical coordinate is key rate.} 
\label{fig:std0}
\centering
\end{figure}
\begin{figure}
\includegraphics[width=0.5\textwidth]{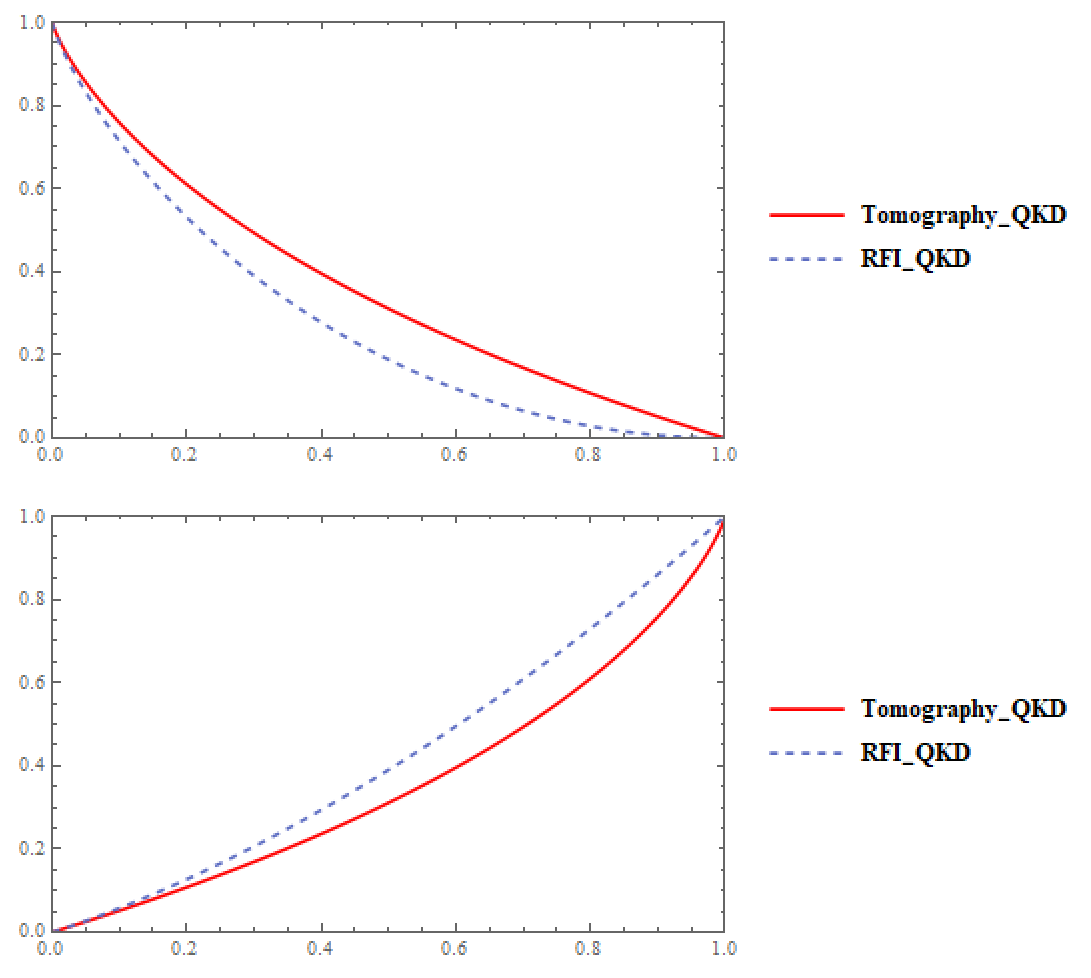}
\caption{The upper panel is the mutual information between Alice and Bob $I(X|Y)$ of tomography-based QKD and RFI-QKD in amplitude damping channel.  The bottom panel is the Holevo quantity between Alice and Eve $\chi(X:E)$. The horizontal coordinate is $p$ and the vertical coordinate is the amount of information in both panels. We can clearly see $I_{\rho_{AB}}(X|Y)>I_{\rho_{AB}'}(X|Y)$ and $\chi_{\rho_{AB}}(X:E)<\chi_{\rho_{AB}'}(X:E)$ when $0<p<1$.} 
\label{fig:std1}
\centering
\end{figure}

The key rate of tomography-based QKD protocol is given by Ref.\cite{PhysRevA.78.042316}:
\begin{equation}
r_{QST}=1+\frac{1}{2} h(p)-h\left(\frac{p}{2}\right)-\frac{1+p}{2} h\left(\frac{1}{1+p}\right)
\end{equation}
We derive the key rate of RFI-QKD protocol, which is:
\begin{equation}
\begin{aligned}
    r_{RFI}=&1+\frac{p}{2}log_2\frac{p}{4}+\frac{2-p-2\sqrt{1-p}}{4}log_2\frac{2-p-2\sqrt{1-p}}{4}\\
    &+\frac{2-p+2\sqrt{1-p}}{4}log_2\frac{2-p+2\sqrt{1-p}}{4}
\end{aligned}
\end{equation}
which we plotted in Figure.4. Clearly we can see the key rate of tomography-based QKD is higher than RFI-QKD.
Next we calculated the mutual information between Alice and Bob $I(X|Y)$ and the Holevo quantity between Alice and Eve $\chi(X:E)$ in the above two protocols.

For tomography-based QKD:
\begin{align*} 
I_{\rho_{AB}}(X|Y)=&\frac{1}{2}+h(\frac{1+p}{2})+\frac{p}{2}log_2\frac{p}{2}+\frac{1-p}{2}log_2\frac{1-p}{2}\\
\chi_{\rho_{AB}}(X:E)=&h(\frac{p}{2})-\frac{1}{2}h(p)
\end{align*}

For RFI-QKD:
\begin{align*} 
I_{\rho_{AB}'}(X|Y)=&1-h(\frac{p}{2})\\
\chi_{\rho_{AB}'}(X:E)=&-h(\frac{p}{2})-\frac{p}{2}log_2\frac{p}{4}\\
&-\frac{2-p-2\sqrt{1-p}}{4}log_2\frac{2-p-2\sqrt{1-p}}{4}\\
&-\frac{2-p+2\sqrt{1-p}}{4}log_2\frac{2-p+2\sqrt{1-p}}{4}
\end{align*}
We then plot them in Figure.5 for comparison. Clearly we can see $I_{\rho_{AB}}(X|Y)>I_{\rho_{AB}'}(X|Y)$ and $\chi_{\rho_{AB}}(X:E)<\chi_{\rho_{AB}'}(X:E)$ when $0<p<1$, which leads to $r_{QST}>r_{RFI}$ in amplitude damping channel for the region $0<p<1$. In a word, the key rate for QST-based QKD is all strictly higher than that for  RFI-QKD for non-trivial values of $p$ ($0<p<1$).
\subsection{Rotation channel}

In this part we will show that for any unitary rotation channel, the key rate of the two protocols are the same. Since any rotation matrix in three dimensional space can be decomposed as $R_{\overrightarrow n}(\theta)=R_y(\alpha_y)R_x(\alpha_x)R_z(\alpha_z)$, where $R_{\overrightarrow n}(\theta)$ denotes a rotation by angle $\theta$ along the $\overrightarrow n$ axis, the general rotation channel can be defined as
\begin{widetext}
\begin{equation}
\left[\begin{array}{l}{\theta_{z}} \\ {\theta_{x}} \\ {\theta_{y}}\end{array}\right] \mapsto\left[\begin{array}{ccc}{cos \alpha_y} & {-sin \alpha_y} & {0} \\ {sin \alpha_y} & {cos \alpha_y} & {0} \\ {0} & {0} & {1}\end{array}\right]
\left[\begin{array}{ccc}{cos \alpha_x} & {0} & {-sin \alpha_x} \\ {0} & {1} & {0} \\ {sin \alpha_x} & {0} & {cos \alpha_x}\end{array}\right]
\left[\begin{array}{ccc}{1} & {0} & {0} \\ {0} & {cos \alpha_z} & {-sin \alpha_z} \\ {0} & {sin \alpha_z} & {cos \alpha_z}\end{array}\right]
\left[\begin{array}{l}{\theta_{z}} \\ {\theta_{x}} \\ {\theta_{y}}\end{array}\right]
\label{2222}
\end{equation}
\end{widetext}
We plug Eq.(\ref{2222}) into the key rate formula in Section.II and V and get

\begin{equation}
r_{QST}=r_{RFI}=1-h(\frac{1+cos\alpha_x cos\alpha_y}{2})
\label{2323}
\end{equation}

The key rates of the tomography-based QKD and RFI-QKD are the same in this situation. 
Next we calculate the mutual information between Alice and Bob $I(X|Y)$ of the Holevo quantity between Alice and Eve $\chi(X:E)$ of the two protocols. We find
\begin{align*} 
I_{\rho_{AB}'}(X|Y)&=I_{\rho_{AB}}(X|Y)=1-h(\frac{1+cos\alpha_x cos\alpha_y}{2})\\
\chi_{\rho_{AB}'}(X:E)&=\chi_{\rho_{AB}}(X:E)=0
\end{align*}

The Holevo quantity $\chi(X:E)$ is zero in both protocols, which means Eve can not gain any information as the channel applies a unitary rotaion to the transmitted state. This leads to $r_{QST}=r_{RFI}$ in Eq.(\ref{2323}).
Take $\alpha_z=0, \alpha_y=0$ for an example, this rotation channel was defined as 
\begin{equation}
\left[\begin{array}{l}{\theta_{z}} \\ {\theta_{x}} \\ {\theta_{y}}\end{array}\right] \mapsto\left[\begin{array}{ccc}{cos \alpha_x} & {0} & {-sin \alpha_x} \\ {0} & {1} & {0} \\ {sin \alpha_x} & {0} & {cos \alpha_x}\end{array}\right]\left[\begin{array}{l}{\theta_{z}} \\ {\theta_{x}} \\ {\theta_{y}}\end{array}\right]
\end{equation}
which means the Bloch vector was rotated by $\alpha$ along the x axis. 
we have 
\begin{equation}
r_{QST}=r_{RFI}=1-h(\frac{1+cos\alpha_x}{2})
\end{equation}
This key rate is plotted in Figure.6.

\begin{figure}
\includegraphics[width=0.5\textwidth]{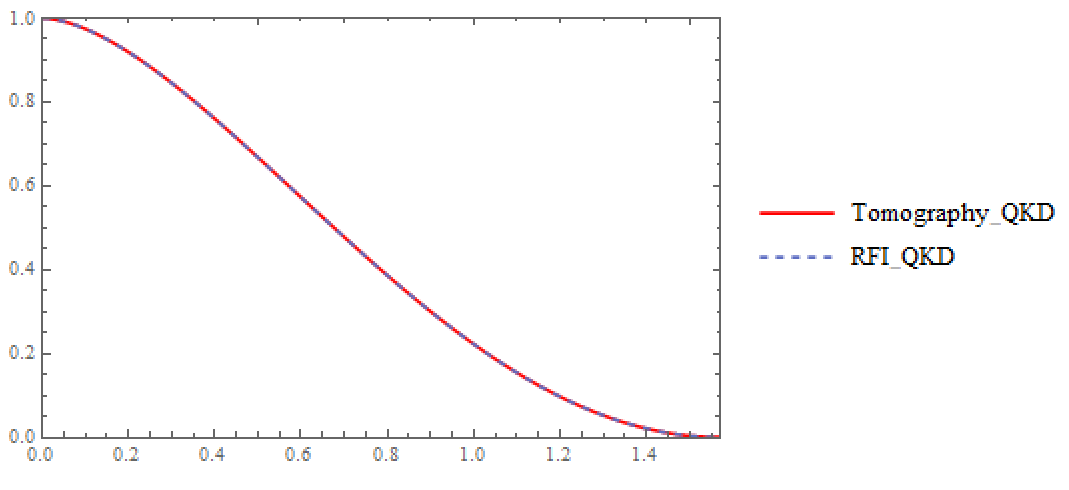}
\caption{The key rates of Tomography-based QKD and RFI-QKD in rotation channel. The horizontal coordinate is the rotation angle $\beta$ and the vertical coordinate is key rate(Direct reconciliation).} 
\label{fig:std0}
\centering
\end{figure}
\subsection{Probabilistic rotation channel}
Assuming the channel between Alice and Bob rotates the state in a probabilistic way. The probability of rotating the Bloch vector along the x axis or y axis are both 1/2. For simplicity we assume the rotation angles are both $\alpha$. This channel can be written as
\begin{equation}
\left[\begin{array}{l}{\theta_{z}} \\ {\theta_{x}} \\ {\theta_{y}}\end{array}\right] \mapsto\frac{1}{2}\left[\begin{array}{ccc}{2cos \alpha} & {-sin \alpha} & {-sin \alpha} \\ {sin \alpha} & {1+cos\alpha} & {0} \\ {sin \alpha} & {0} & {1+cos \alpha}\end{array}\right]\left[\begin{array}{l}{\theta_{z}} \\ {\theta_{x}} \\ {\theta_{y}}\end{array}\right]
\end{equation}
In this situation, the entropy of the joint state of Alice and Bob is $S(\rho_{AB})=h((1-cos\alpha)/4)$. The key rate of tomography-based QKD and RFI-QKD are calculated as
\begin{equation}
r_{QST}=1-h(\frac{1-cos\alpha}{4})-h(\frac{1-cos\alpha}{2})+h(\frac{1+\sqrt{(1+cos^2 \alpha)/2}}{2})
\end{equation}
and \begin{equation}
    r_{RFI}=\frac{1+cos \alpha}{2}-h(\frac{1+cos\alpha}{2})
\end{equation}
which we plotted in Figure.7. Then we calculated the mutual information $I(X|Y)$ and the Holevo quantity $\chi(X:E)$ of the two protocols.

For tomography-based QKD:
\begin{align*} 
I_{\rho_{AB}}(X|Y)=&1-h(\frac{1-cos \alpha}{2})\\
\chi_{\rho_{AB}}(X:E)=&h(\frac{1-cos \alpha}{4})-h(\frac{1-\sqrt{(1+cos^2\alpha)/2}}{2})
\end{align*}

For RFI-QKD:
\begin{align*} 
I_{\rho_{AB}'}(X|Y)=&1-h(\frac{1-cos \alpha}{2})\\
\chi_{\rho_{AB}'}(X:E)=&\frac{1-cos \alpha}{2}
\end{align*}
We then plot them in Figure.8 for comparison. We can see that $I_{\rho_{AB}}(X|Y)=I_{\rho_{AB}'}(X|Y)$ and $\chi_{\rho_{AB}}(X:E)<\chi_{\rho_{AB}'}(X:E)$ when $0<\alpha<\pi$, which lead to $r_{QST}>r_{RFI}$ in probabilistic rotation channel for $0<\alpha<\pi$.
\begin{figure}
\includegraphics[width=0.5\textwidth]{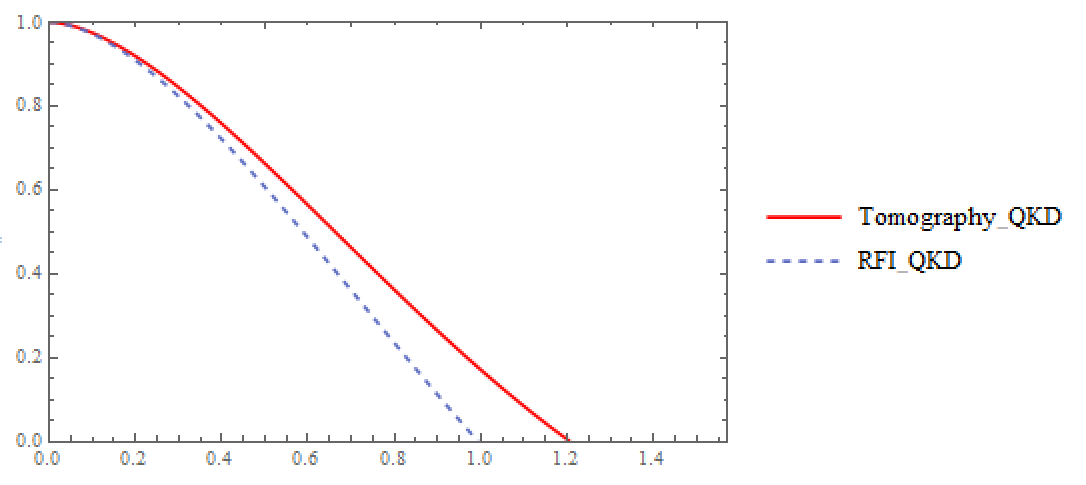}
\caption{The key rates of Tomography-based QKD and RFI-QKD in probabilistic rotation channel. The horizontal coordinate is the rotation angle $\alpha$ and the vertical coordinate is key rate(Direct reconciliation).} 
\label{fig:std3}
\centering
\end{figure}
\begin{figure}
\includegraphics[width=0.5\textwidth]{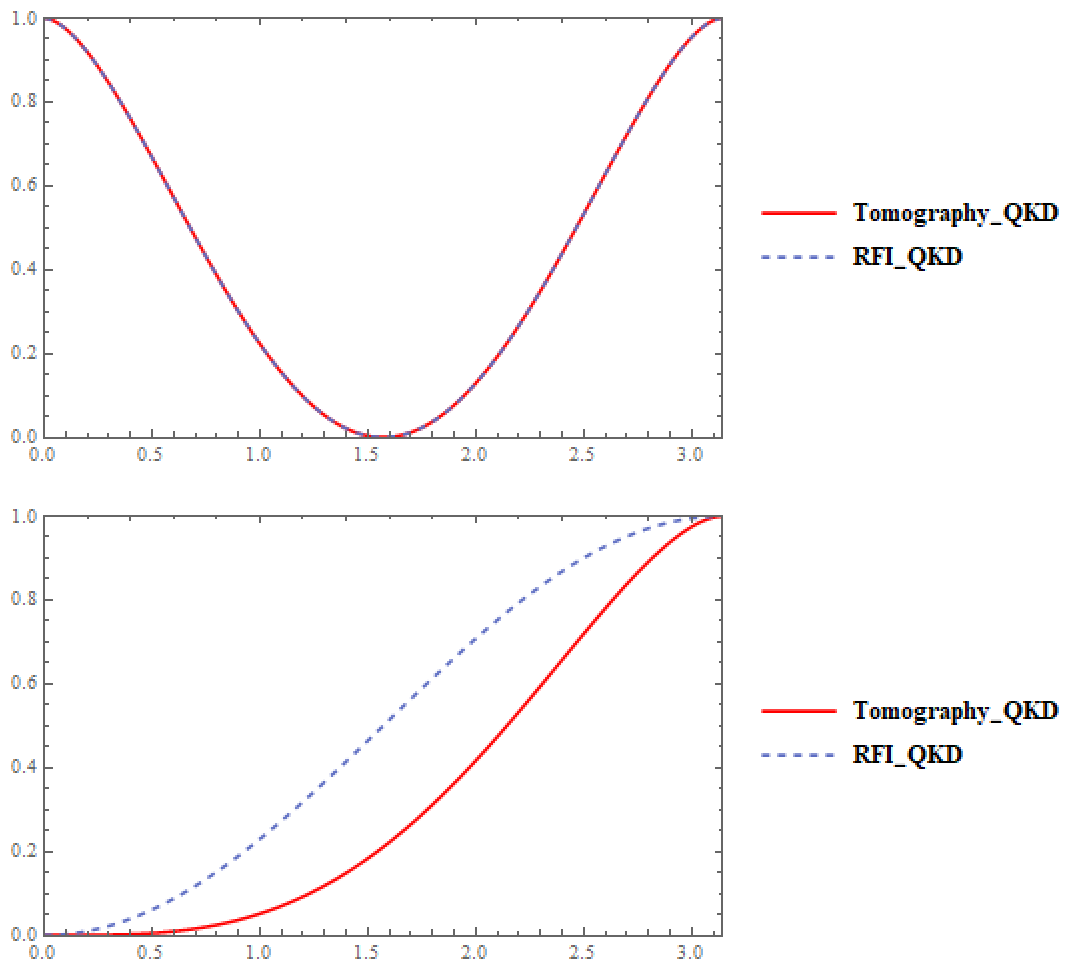}
\caption{The upper panel is the mutual information between Alice and Bob $I(X|Y)$ of Tomography-based QKD and RFI-QKD in probabilistic rotating channel.  The bottom panel is the Holevo quantity between Alice and Eve $\chi(X:E)$. The horizontal coordinate is $\alpha$ and the vertical coordinate is the amount of information in both panels. We can clearly see $I_{\rho_{AB}}(X|Y)=I_{\rho_{AB}'}(X|Y)$ and $\chi_{\rho_{AB}}(X:E)<\chi_{\rho_{AB}'}(X:E)$ when $0<\alpha<\pi$.} 
\label{fig:std0}
\centering
\end{figure}

\section{Comparison in experiment}
There are multiple experimental implementation of RFI-QKD\cite{Wabnig_2013,sc,PhysRevA.95.032309,PhysRevLett.112.130501}. However, no QST-QKD experiment has been done yet. The difference of the two protocol is that QST-QKD includes full tomography of the quantum channel while RFI-QKD does not. So if we want to investigate the differences of the two protocols in experimental implementation, we just have to find their differences in dealing with the defects in real life quantum channels(optical fibre).

\textbf{Photon loss}

Usually the channel loss is related to the transmission distance by a loss coefficient $\beta_c$ in dB/km. The transmittance $\eta$ is given by 
\begin{equation}
    \eta=\eta_B 10^{-\frac{\beta_c l}{10}}
\end{equation}
where $\eta_B$ denotes the transmittance on Bob's side. 

The QST-QKD and RFI-QKD protocols are both postprocessing protocols. So the key rate of both protocols are not affected by photon loss. The transmittance of the quantum channel is reduced by $\eta$.

\textbf{Polarization misalignment}

Many commonly used fibres are made of birefringent crystals. It is difficult to maintain the polarization in these fibres. As we have analyzed in Section VII.B, RFI-QKD and QST-QKD protocols have the same key rate if the quantum channel applies a unitary rotation to the transmitted qubit.

\textbf{Polarization stability}

Randoms changes of birefringence in optical fibre due to enviromental variations lead to slow random unitary transformations. This random unitary transformations can be described as:
$R(t)=R_y(\gamma+\Delta\gamma(t))R_x(\beta+\Delta\beta(t))R_z(\alpha+\Delta\alpha(t))$, where
\begin{equation}
R_{y}(\gamma+\Delta\gamma(t))=\left[\begin{array}{ccc}{cos (\gamma+\Delta\gamma(t))} & {-sin (\gamma+\Delta\gamma(t))} & {0} \\ {sin (\gamma+\Delta\gamma(t))} & {cos (\gamma+\Delta\gamma(t))} & {0} \\ {0} & {0} & {1}\end{array}\right]
\end{equation}
\begin{equation}
    R_{x}(\beta+\Delta\beta(t))=\left[\begin{array}{ccc}{cos (\beta+\Delta\beta(t))} & {0} & {-sin (\beta+\Delta\beta(t))} \\ {0} & {1} & {0} \\ {sin (\beta+\Delta\beta(t))} & {0} & {cos (\beta+\Delta\beta(t))}\end{array}\right]
\end{equation}
\begin{equation}
    R_z(\alpha+\Delta\alpha(t))=\left[\begin{array}{ccc}{1} & {0} & {0} \\ {0} & {cos (\alpha+\Delta\alpha(t))} & {-sin (\alpha+\Delta\alpha(t))} \\ {0} & {sin (\alpha+\Delta\alpha(t))} & {cos( \alpha+\Delta\alpha(t))}\end{array}\right]
\end{equation}
The average transformation is
\begin{equation}
    \bar R=\frac{1}{T}\int_0^T R_y(\gamma+\Delta\gamma(t))R_x(\beta+\Delta\beta(t))R_z(\alpha+\Delta\alpha(t)) dt
\end{equation}
This is a generalized version of probabilistic rotation channel, which we have analyzed in part 4.3. We guess that $r_{QST}>r_{RFI}$ always holds. 

We take $\alpha=\beta=\gamma=\pi/6, \Delta\alpha\sim N(0,(\pi/12)^2)(\Delta\alpha,\Delta\beta,\Delta\gamma, i.i.d.)$ for an example. In this case, $r_{QST}=0.412>r_{RFI}=0.367$.

\textbf{Polarization mode dispersion(PMD)}

For polarization-entangled photon pairs transmitted in optical fibre, PMD is the chief polarization decoherence mechanism\cite{PhysRevLett.106.080404}. Alice prepare the entangled photon pair before sending one of them to Bob. The initial state of the photon pair is:
\begin{equation}
|\psi\rangle=\iint d t_{A} d t_{B} f\left(t_{A}-t_{B}\right)\left|t_{A}, t_{B}\right\rangle \otimes \frac{\left|u_A u_B\right\rangle+ \left|u'_A u'_B\right\rangle}{\sqrt{2}}
\end{equation}

where $u_A(u_B)$ and $u'_A(u'_B)$ denote the orthonormal basis of the polarization state of photon A(B). We then represent the polarization dependent part in the basis of principal states of the PMD in the two arms.
\begin{equation}
\begin{aligned}
\left|\psi_{p}\right\rangle =&(\left|u_A u_B\right\rangle+ \left|u'_A u'_B\right\rangle)/\sqrt{2}\\=&\eta_{1}\left(\left|{s}_{A}, {s}_{B}\right\rangle+ e^{i \tilde{\alpha}_{1}}\left|{s}_{A}^{\prime}, {s}_{B}^{\prime}\right\rangle\right) / \sqrt{2} \\
&+\eta_{2}\left(\left|{s}_{A}, {s}_{B}^{\prime}\right\rangle- e^{i \tilde{\alpha}_{2}}\left|{s}_{A}^{\prime}, {s}_{B}\right\rangle\right) / \sqrt{2}
\end{aligned}
\end{equation}

where $s_A(s_B)$ and $s'_A(s'_B)$ denote the principle states of polarization along the path of photon A(B). The coefficients are given by
\begin{equation}
\begin{aligned}
&\eta_{1}=\left({s}_{A} \cdot {u}_{A}\right)\left({s}_{B} \cdot {u}_{B}\right)+\left({s}_{A} \cdot {u}_{A}^{\prime}\right)\left({s}_{B} \cdot {u}_{B}^{\prime}\right)\\
&\eta_{2}=\left({s}_{A} \cdot {u}_{A}\right)\left({s}_{B}^{\prime} \cdot {u}_{B}\right)+\left({s}_{A} \cdot {u}_{A}^{\prime}\right)\left({s}_{B}^{\prime} \cdot {u}_{B}^{\prime}\right)
\end{aligned}
\end{equation}
The phase factor $\tilde{\alpha}_i$ is defined through  $\eta_{i}=\left|\eta_{i}\right| \exp (-i\tilde{\alpha}_{i}/2)$. The final state of two entangled photons after propagating through the media is
\begin{equation}
\begin{aligned}
\left|\psi_{\text {out }}\right\rangle &=\frac{\eta_{1}}{\sqrt{2}}\left|f\left(t_{A}-t_{B}-\frac{\tau_{A}-\tau_{B}}{2}\right)\right\rangle \otimes\left|\underline{s}_{A}, \underline{s}_{B}\right\rangle \\
&+\frac{\eta_{2}}{\sqrt{2}}\left|f\left(t_{A}-t_{B}-\frac{\tau_{A}+\tau_{B}}{2}\right)\right\rangle \otimes\left|\underline{s}_{A}, \underline{s}_{B}^{\prime}\right\rangle \\
&-\frac{\eta_{2}^{*} e^{i \tilde{\alpha}_{2}}}{\sqrt{2}}\left|f\left(t_{A}-t_{B}+\frac{\tau_{A}+\tau_{B}}{2}\right)\right\rangle \otimes\left|\underline{s}_{A}^{\prime}, \underline{s}_{B}\right\rangle \\
&+\frac{\eta_{1}^{*} e^{i \tilde{\alpha}_{1}}}{\sqrt{2}}\left|f\left(t_{A}-t_{B}+\frac{\tau_{A}-\tau_{B}}{2}\right)\right\rangle \otimes\left|\underline{s}_{A}^{\prime}, \underline{s}_{B}^{\prime}\right\rangle
\end{aligned}
\end{equation}
The analysis of PMD effect above is given in Ref.\cite{PhysRevLett.106.080404}. Now we consider how PMD effect affects the key rate of quantum communication.

The density matrix of the final state is
\begin{equation}
\rho_{AB}=\iint d t_{A} d t_{B} |\psi_{out}\rangle \langle \psi_{out}|\label{den}
\end{equation}

Note that the QST-QKD and RFI-QKD protocol have same key rate formula but for different density matrix $\rho_{AB}$ and $\rho_{AB}'$. For the density matrix in Eq.(\ref{den}), generally $\rho_{AB}\neq\rho_{AB}'$, which means the key rate for the two protocols are different. The proof of $\rho_{AB}\neq\rho_{AB}'$ is not given here because the explicit form of $\rho_{AB}$ is complicated. 

Take a simple case for example. We assume that $u_A$ and $u_B$ are horizontal polarized state, denoted by $|0\rangle$, and $u'_A$ and $u'_B$ are vertical polarized state, denoted by $|1\rangle$. Alice holds her photon and sends another  photon to Bob, i.e. $\tau_A=0$. The principle state of polarization in the optical fibre is slightly shifted from Alice's basis by an angle $\beta$, which means:
\begin{equation}
\begin{aligned}
    &|s_A\rangle=|0\rangle\\
    &|s'_A\rangle=|1\rangle\\
    &|s_B\rangle=cos \beta|0\rangle+sin \beta |1\rangle\\
    &|s'_B\rangle=-sin \beta|0\rangle+cos \beta |1\rangle
    \end{aligned}
\end{equation}


           
We define an autocorrelation function $R(\tau_B)$ as $R(\tau_B)=\int dt f^*(t)f(t+\tau_B)$ where $\tau_B$ is the time difference between signals propagating in different polarization modes. We calculate and depict the key rates of RFI-QKD and QST-QKD protocols as a function of $\beta$ in Fig.6. The explicit form of the key rates are complicated so we didn't present them here. The result shows that usually $r_{QST-QKD}$ is higher than $r_{RFI-QKD}$, especially when R is small, i.e. the time difference is big.

\begin{figure*}
\centering
\includegraphics[width= \textwidth]{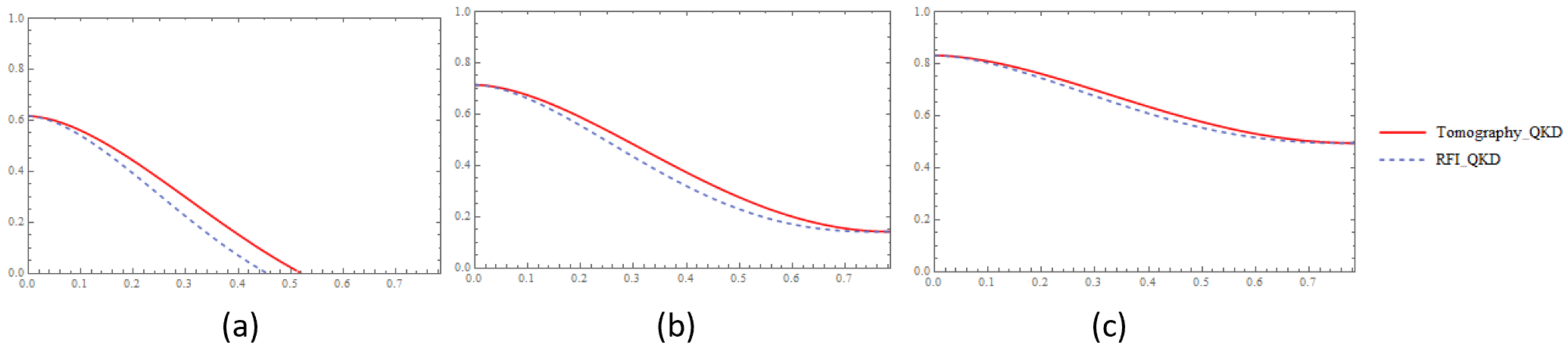}
\caption{The key rate of Tomography-based QKD and RFI-QKD as a function of $\beta$ in quantum channel with PMD effect. The horizontal coordinate is $\beta$ and the vertical coordinate is key rate. In panel.(a), $R=0.85$. In panel.(b), $R=0.9$. In panel.(c), $R=0.95$.} 
\label{fig:std5}
\centering
\end{figure*}

\textbf{Polarization dependent loss(PDL)}\cite{npj,Kirby:19}

We assume the transmmittance of $|0\rangle$($|1\rangle$) is $\eta^0$($\eta^1$) in the quantum channel with PDL effect. In the case of that input state is Bell state, the output state is
\begin{equation}
|\Psi_{final}\rangle=(\sqrt{\eta^0}\left|00\right\rangle+ \sqrt{\eta^1}\left|11\right\rangle)/\sqrt{\eta^0+\eta^1}
\end{equation}
The key rate of QST-QKD and RFI-QKD protocols are
\begin{equation}
    r_{QST}=r_{RFI}=h(\eta^0/(\eta^0+\eta^1))
\end{equation}
The two protocols have the same key rate in quantum channels with PDL effect.
\section{Conclusion}
In this paper we extend the tomography-based QKD to high dimensional cases. We show how to perform full channel tomography using Alice and Bob's measurement results. Compared to the traditional $(d+1)$-basis protocol, the high-dimensional tomography-based protocol can achieve higher key rate. We demonstrate that by both theoretical analysis and applying our protocol to previous experimental data.

Then we compare the tomography-based QKD protocol to RFI-QKD protocol. By numerical calculating the two protocols in different channels, we show that the tomography-based QKD protocol is a better choice for real-life QKD implementation for it can achieve higher key rate.
\section{Acknowledgement}
We acknowledge discussions with colleagues including Eli Bourassa and Wenyuan Wang. We thank funding support from NSERC, CFI, ORF, Huawei Technologies Canada, MITACS, US Office of Naval Research, Royal Bank of Canada and the University of Hong Kong start-up grant.




\section{Appendix}
Here we give a rigorous proof of Eq.(\ref{ineq}) by showing  $I_{\rho_{AB}}(X|Y)\geq I_{\rho_{AB}'}(X|Y)$ and $\chi_{\rho_{AB}}(X:E)\leq\chi_{\rho_{AB}'}(X:E)$.

\textbf{Proof for $I_{\rho_{AB}}(X|Y)\geq I_{\rho_{AB}'}(X|Y)$:}

We have 
\begin{equation}
    I_{\rho_{AB}}(X|Y)=h(\frac{1+t_z}{2})-\frac{1}{2}h(\frac{1-R_{zz}+t_z}{2})-\frac{1}{2}h(\frac{1+R_{zz}+t_z}{2})
\end{equation}
\begin{equation}
    I_{\rho_{AB}'}(X|Y)=1-h(\frac{1-R_{zz}}{2})
\end{equation}
We can rewrite Eq.(30) as
\begin{equation}
    I_{\rho_{AB}'}(X|Y)=h(\frac{1}{2})-\frac{1}{2}h(\frac{1-R_{zz}}{2})-\frac{1}{2}h(\frac{1-R_{zz}}{2})
\end{equation}
Define a function $f(x)$:
\begin{equation}
     f(x)=h(\frac{1+x}{2})-\frac{1}{2}h(\frac{1-R_{zz}+x}{2})-\frac{1}{2}h(\frac{1+R_{zz}+x}{2})
\end{equation}
It is easy to prove $f(x)$ is a decreasing function when $x<0$ and a increasing function when $x>0$. $x=0$ is the minimum point of $f(x)$, so we have $f(0)\leq f(t_z)$, which is $I_{\rho_{AB}}(X|Y)\geq I_{\rho_{AB}'}(X|Y)$.

\textbf{Proof for $\chi_{\rho_{AB}}(X:E)\leq\chi_{\rho_{AB}'}(X:E)$:}

We have
\begin{equation}\label{s1}
    \chi_{\rho_{AB}}(X:E)=S(\rho_{AB})-\frac{1}{2}S(_A\langle 0|\rho_{AB}|0 \rangle_A)-\frac{1}{2}S(_A\langle 1|\rho_{AB}|1 \rangle_A)
\end{equation}
\begin{equation}\label{s2}
    \chi_{\rho_{AB}'}(X:E)=S(\rho_{AB}')-\frac{1}{2}S(_A\langle 0|\rho_{AB}'|0 \rangle_A)-\frac{1}{2}S(_A\langle 1|\rho_{AB}'|1 \rangle_A)
\end{equation}
The conditional von Neumann entropy $H_{\rho}(X|E)$ can be expressed as
\begin{align*}
H_{\rho}(X|E)=&H(\rho_{XE})-H(\rho_E)\\=&H(X)+\frac{1}{2}H(\mathcal{E}_B(|0\rangle\langle 0|))+\frac{1}{2}H(\mathcal{E}_B(|1\rangle\langle 1|))-H(\rho_{AB})
\end{align*}
So we have $\chi_{\rho_{AB}}(X:E)=H_{\rho_{AB}}(X|E)-1$, $\chi_{\rho_{AB}'}(X:E)=H_{\rho_{AB}'}(X|E)-1$.

We use the Lemma 2 in Appendix A in Ref.\cite{PhysRevA.78.042316} and get
\begin{equation}
H_{\tilde{\rho}_{AB}}(X|E)\leq H_{\rho_{AB}'}(X|E)
\end{equation}
\begin{equation}
H_{{\rho}_{AB}}(X|E)\leq H_{\tilde{\rho}_{AB}}(X|E)
\end{equation}
Hence we have $H_{{\rho}_{AB}}(X|E)\leq H_{\rho_{AB}'}(X|E)$, which is $\chi_{\rho_{AB}}(X:E)\leq\chi_{\rho_{AB}'}(X:E)$.


\textbf{Proof of $\chi_{\rho'}\leq\chi_{\rho}$}
For simplicity we only prove  the $d=2$ case here. For higher dimensions, the proof is similar.

We use the Lemma 2 in Appendix A in Ref.[3].

Lemma:
For two channels $\mathcal{E}_B^1$ and $\mathcal{E}_B^2$ and a probabilistically mixed channel $\mathcal{E}_{B}^{\prime}:=\lambda \mathcal{E}_{B}^{1}+(1-\lambda) \mathcal{E}_{B}^{2}$, Eve's ambiguity is convex, i.e. we have

\begin{equation}\chi_{\rho^{\prime}} \leqslant \lambda \chi_{\rho^{1}}+(1-\lambda) \chi_{\rho^{2}}\end{equation}

Define $\sigma_1=diag(1,1,1,-1),\sigma_2=diag(1,-1,1,1)$, we have
\begin{align}
    \tilde{\rho}&=\frac{1}{2}(\rho+Z_AZ_B\rho Z_AZ_B) \\
    \rho'&=\frac{1}{2}(\sigma_1\rho\sigma_1+\sigma_2\rho \sigma_2)
\end{align}

where $Z_A$ and $Z_B$ are Pauli $Z$ operators. Using the lemma mentioned above, we get
\begin{align}
    \chi_{\rho^{\prime}}& \leqslant \chi_{\tilde{\rho}} \\
    \chi_{\tilde{\rho}}& \leqslant \chi_{\rho}
\end{align}

So $\chi_{\rho'}\leq\chi_{\rho}$.

\textbf{Amplitude Damping Channel}

In previous sections we show that if Alice and Bob's joint state is not Bell-diagonal, the tomography-based protocol gives a higher key rate than $(d+1)$-basis protocol. Here we give an example of Alice and Bob's joint state in qubit amplitude damping channel. The channel is described by Eq.(\ref{1919}) in Section VII.A. The joint state of Alice and Bob in computational basis is
\begin{equation}
\rho_{AB}=\frac{1}{4}\left[\begin{array}{cccc}{2} & {0} & {0} & {2\sqrt{1-p}} \\ {0} & {0} & {0}&{0} \\ {0} & {0} & {2p}&{0} \\ {2\sqrt{1-p}} & {0} & {0} & {2-2p}\end{array}\right]
\end{equation}
The density matrix in Bell basis is

\begin{equation}
\rho_{AB}=\frac{1}{4}\left[\begin{array}{cccc}{2-p+2\sqrt{1-p}} & {0} & {0} & {p} \\ {0} & {p} & {-p}&{0} \\ {0} & {-p} & {p}&{0} \\ {p} & {0} & {0} & {2-p-2\sqrt{1-p}}\end{array}\right]
\end{equation}

\begin{figure}
\includegraphics[width=0.5 \textwidth]{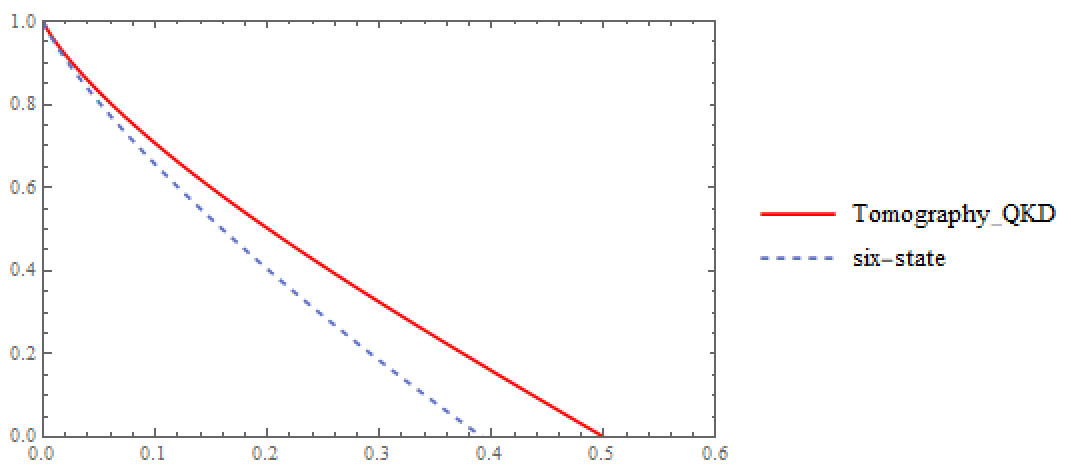}
\caption{The key rate of tomography-based QKD protocol and six-state protocol in amplitude damping channel. The vertical axis is $p$  and the horizontal axis is the secret key rate.} 
\label{1010}
\centering
\end{figure}
We can clearly see that the density matrix is not diagonal in Bell basis, that's why the tomography-based QKD protocol gives a higher key rate than six-state protocol in qubit amplitude damping channel, as we have shown in Figure.\ref{1010}.

\bibliography{apssamp}

\nocite{*}


\end{document}